\documentclass{aastex61}
\usepackage{graphicx,natbib}
\usepackage[percent]{overpic}
\usepackage{hyperref}

\shorttitle{Prominence carrying coronal flux rope}
\shortauthors{Fan et al.}

\begin{document}

\title{The eruption of a prominence carrying coronal flux rope:
forward synthesis of the magnetic field strength measurement by the
COronal Solar Magnetism Observatory Large Coronagraph}

\correspondingauthor{Yuhong Fan}
\email{yfan@ucar.edu}

\author[0000-0003-1027-0795]{Yuhong Fan}
\affil{High Altitude Observatory, National Center for Atmospheric Research, 3080 Center Green Drive, Boulder, CO 80301, USA}

\author{Sarah Gibson}
\affil{High Altitude Observatory, National Center for Atmospheric Research, 3080 Center Green Drive, Boulder, CO 80301, USA}

\author{Steve Tomczyk}
\affil{High Altitude Observatory, National Center for Atmospheric Research, 3080 Center Green Drive, Boulder, CO 80301, USA}

\begin{abstract}
From a magnetohydrodynamic (MHD) simulation of the eruption of a prominence hosting coronal flux rope,
we carry out forward synthesis of the circular polarization signal (Stokes V signal) of
the FeXIII emission line at 1074.7 nm produced by the MHD model as measured by the proposed
COronal Solar Magnetism Observatory (COSMO) Large Coronagraph (LC) and infer the line-of-sight
magnetic field $B_{\rm LOS}$ above the limb.
With an aperture of 150 cm,
integration time of 12 min,
and a resolution of 12 arcsec, the LC can
measure a significant $B_{\rm LOS}$ with sufficient signal to noise level, from the simulated flux
rope viewed nearly along its axis with a peak axial field strength of about 10 G. The measured $B_{\rm LOS}$ is found to
relate well with the axial field strength of the flux rope for the height range of the prominence,
and can discern the increase with height of the magnetic field strength in that
height range that is a definitive signature of the concave upturning dipped field supporting the prominence.
The measurement can also detect an outward moving $B_{\rm LOS}$ due to the slow
rise of the flux rope as it develops the kink instability, during the phase when its rise speed is still below
about 41 km/s and up to a height of about 1.3 solar radii. These results suggest that the COSMO LC has great
potential in providing quantitative information about the magnetic field structure of CME precursors
(e.g. the prominence cavities) and their early evolution for the onset of eruption.
\end{abstract}


\section{Introduction}

Coronal mass ejections (CMEs) often originate from regions in the solar corona where
there are prominences or filaments, which are elongated large-scale structures of
cool and dense plasma suspended in the much hotter and rarefied solar corona 
supported by the magnetic fields \citep[e.g.][]{Webb:Hundhausen:1987,Gibson:2018}.
Features observed surrounding the prominences such as cavities \citep[e.g.][]{Gibson:2015}
and hot shrouds \citep[e.g.][]{Hudson:etal:1999,Habbal:etal:2010} represent the
coronal environment of the magnetic field structures hosting the prominence.
A quantitative measurement of the magnetic field strength and spatial properties of
such magnetic structures in the corona would significantly advance our
understanding of the physical conditions and mechanisms for the development of CMEs.
Direct measurement of the coronal magnetic field strength has been
rare \citep[e.g.][]{Lin:etal:2000} and extremely difficult because of the weakness of
the coronal magnetic fields and the extremely faint solar corona emission.
The optically thin nature of the plasma is also one of the principal reasons for a
difficult magnetic field measurement in the corona.

The COronal Solar Magnetism Observatory (COSMO), is a proposed synoptic facility designed
to measure magnetic fields and plasma properties in the large-scale solar atmosphere
\citep{Tomczyk:etal:2016}. Among the suite of three instruments proposed for COSMO is
the Large Coronagraph (LC) with a 1.5m aperture to measure the magnetic field,
temperature, density, and dynamics of the corona above the solar limb.
The COSMO LC measures the line-of-sight (LOS) strength of coronal  magnetic fields
directly through the Zeeman effect observed in the circular polarization (Stokes V
profile) of coronal forbidden emission lines, including the FeXIII emission line at
1074.7 nm.  The theoretical formalism for calculating the Stokes profiles of
forbidden emission lines such as the FeXIII 1074.7 nm line under coronal
conditions has been developed in \citet{Casini:Judge:1999}. A Fortran-77
program, the Coronal Line Emission (CLE) code, that implements the formalism
and synthesize the Stokes profiles
given the coronal plasma and magnetic field conditions along the observed
line-of-sight is described in \citet{Judge:Casini:2001}.
Using this code, \citet{Judge:etal:2006} carried out the first forward
calculations of the
Stokes signals of several coronal emission lines produced by a
global, axisymmetric current-carrying magnetic structure embedded in a
background isothermal corona to study the signatures of non-potential
coronal magnetic fields.
\citet{Gibson:etal:2016} has incorporated the CLE code into the SolarSoft
FORWARD package and presented forward synthesis of the Stokes signals
based on several coronal MHD models.

In this paper, we carry out forward synthesis of the
Stokes V signal of the 
FeXIII 1074.7 nm emission line produced by an
MHD simulation of a prominence carrying coronal flux rope \citep{Fan:2017},
as would be observed by the COSMO LC.
We study the feasibility of measuring the LOS magnetic field of
the prominence flux rope above the limb, viewed nearly along its
length, from the Stokes V signal measured by the COSMO LC
given the error estimation \citep{Tomczyk:2015}.
Although the MHD simulation is still highly simplified in its
treatment of the thermodynamics, using an empirical coronal
heating that depends on height, it does form a prominence condensation
supported by the flux rope due to the radiative instability during
the quasi-static phase and later develops a prominence
eruption as the flux rope erupts due to development of the kink
instability \citep{Fan:2017}.
Thus we use this MHD model to carry out the forward calculation to
get an initial examination of the capability of COSMO LC for
quantitatively inferring the magnetic field strength of the
observed longitudinally extended prominence-cavity systems
\citep[e.g.][]{Gibson:2015},
during both the stable phase as well as the initiation of eruption.

\section{The MHD Model and the Forward Synthesis \label{sec:model}}
\begin{figure}[htb!]
\centering
\includegraphics[width=1.\textwidth]{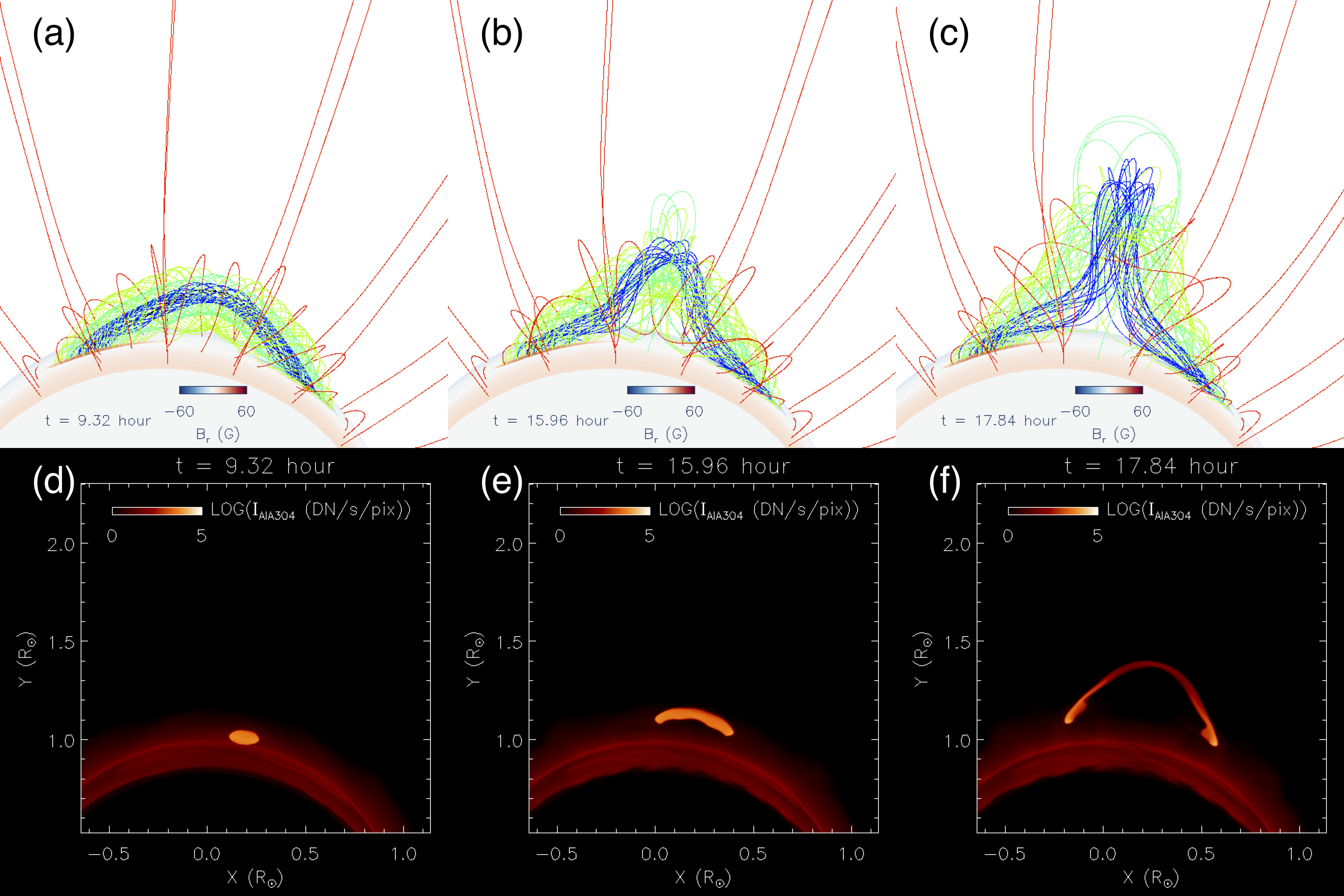}
\caption{Snapshots of the 3D magnetic field lines viewed from the
broad side of the flux rope (upper panels) and the corresponding
synthetic SDO/AIA 304 channel emission images (lower panels) showing
the prominence condensation carried by the flux rope, produced
by the simulation ``WS-L'' case described in F17. A movie of
the evolution is available in the online version of
the paper.}
\label{fig:fig_fdl_aia304_ev}
\end{figure}

The MHD model we use for the forward modeling of the COSMO/LC
observables is the simulation of a longitudinally extended prominence
carrying coronal flux rope
described in the ``WS-L'' case in \citet[][here after F17]{Fan:2017}.
The MHD numerical model, the simulation setup, and the resulting evolution
of the prominence hosting coronal flux rope are described in detail
in F17.  In this case, a twisted magnetic flux rope is driven
quasi-statically into a pre-existing coronal streamer by an imposed
magnetic flux emergence at the lower boundary.
The total field line twist in the emerged flux rope reaches
about 1.83 winds when the emergence is stopped, exceeding the critical limit
for the onset of the kink instability. Subsequently the flux rope becomes
kinked but remains confined, undergoing a slow, quasi-static rise phase,
until it reaches a certain height where it can no longer be confined and
develops an ejective eruption.  During the quasi-static rise phase,
cool prominence condensations are found to form in the dips of the twisted
field lines due to the radiative instability driven by the optically
thin radiative cooling, and as the flux rope erupts, a prominence
eruption is produced.
Figure \ref{fig:fig_fdl_aia304_ev} shows a few snapshots of the evolution
of the 3D magnetic field lines (upper panels) and the prominence the
flux rope carries as visualized by synthetic SDO/AIA 304 channel
emission images (lower panels) viewed from the broad side of the flux
rope.
The way the SDO/AIA 304 channel emission images are computed is described
in F17 (see eq. (23) in that paper).
As can be seen from the Figure, the prominence condensation forms
in the middle dipped portions of the flux rope field lines. As
described in F17, the prominence carrying flux rope undergoes a
quasi-static slow rise phase as it develops the kink instability,
until roughly $t = 17.5$ hr, when it develops a ``hernia-like''
ejective eruption.

Given the simulation data of the magnetic field, plasma density, 
temperature, and line of sight velocity, we synthesize the observed
Stokes I and V profiles of the
forbidden emission line of FeXIII (1074.7nm) by
computing and integrating the emission coefficients
along individual lines of sight (LOS) through the simulation domain
with the CLE code.
A line-of-sight (LOS) magnetic field $B_{\rm LOS}$ can be inferred
from the observed Stokes V circular polarization profile $V_{\lambda}$
of the emission line assuming the magnetograph formula
\citep[e.g.][]{Tomczyk:2015}:
\begin{equation}
V_{\lambda} = -k B_{\rm LOS} \frac{\partial I_{\lambda}}{\partial \lambda},
\end{equation}
where $\lambda$ denotes the wavelength, $k=8.1 \times 10^{-6} {\rm nm}/{\rm G}$
denotes the Zeeman sensitivity for the FeXIII line at 1074.7 nm, and
$I_{\lambda}$ denotes the line intensity profile.  Assuming the emission
line profile is approximately Gaussian:
\begin{equation}
I_{\lambda}=I_0 \exp \left( - \frac{(\lambda - \lambda_0 )^2}{{\Gamma}^2}
\right ) ,
\end{equation}
where $\lambda_0$ is the line-center wavelength (where $I_{\lambda}$ peaks)
and $\Gamma$ is the e-folding line width, it can be shown that:
\begin{equation}
B_{\rm LOS} = \frac{\sqrt{\pi}}{2} \frac{\Gamma}{k} \frac{V}{I},
\label{eqn:blos}
\end{equation}
where
\begin{equation}
V = \int V_{\lambda} \, {\rm sgn} (\lambda - \lambda_0) \, d \lambda
\label{eqn:V}
\end{equation}
and
\begin{equation}
I = \int I_{\lambda} \, d \lambda
\label{eqn:I}
\end{equation}
are the wave-length integrated Stokes $V$ and $I$ signals observed.
One can estimate an error or uncertainty of the
$B_{\rm LOS}$ inferred from the observed $V/I$ signals of the Fe XIII line due to photon noise
\citep{Tomczyk:2015, Lin:2017}:
\begin{equation}
\sigma_{B} = \frac{\sqrt{\pi}}{2} \frac{\Gamma}{k}
\frac{1}{\epsilon_V \sqrt{\Phi_I}}
\left( 1 + 2 \frac{\Phi_{sc}}{\Phi_I} \right )^{1/2} \; ({\rm G}),
\label{eqn:noise}
\end{equation}
where $\sigma_{B}$ is the error of $B_{\rm LOS}$,
$\Phi_I$ denotes the total number of detected photons in the line:
\begin{equation}
\Phi_I = 1.01 \times 10^6 I_0 \, \sqrt{\pi} \, \Gamma \, \Delta t \, \epsilon \,
(\Delta x)^2 \, D^2 \; ({\rm photons}),
\label{eqn:photons}
\end{equation}
and $\Phi_{sc}$ is the background scattered light photons over the same
wavelength range.  In the above $I_0$ is the line
center intensity in units of parts per million (ppm) of the solar disk
center intensity, $\Gamma$ is the observed line width in units of nm,
$\Delta t$ is
the integration time in seconds, $\epsilon$ is the system efficiency,
$\epsilon_V$ is the modulation efficiency for Stokes V,
$\Delta x$ is the size of the spatial resolution element in arcsec,
$D$ is the telescope diameter in meters.
$\Phi_{sc}$ is computed the same way as $\Phi_I$ but with $I_0$ in
equation (\ref{eqn:photons}) replaced with the background corona brightness
in ppm, which we assume to be 5 ppm for the noise estimate.
This level of the background coronal brightness is adopted because it is
achieved with the COSMO K-Coronagraph (K-Cor) and the Coronal Multi-Channel
Polarimeter (CoMP) instruments at the Mauna Loa Solar Observatory (MLSO).
We have used
$\Delta t = 714 $ sec (about 12 min),
$\epsilon=0.05$, $\epsilon_V=1/\sqrt{3}$, 
$\Delta x = 12 $ arcsec, $D = 1.5 $ m for the observation with the COSMO LC.
Although the LC has planned to have the capability to observe at a resolution
of $\Delta x = 2$ arcsec,
we here use a larger measurement pixel size of 12 arcsec to
gain signal to noise, and to accommodate a
reasonably short integration time for detecting
the early phase of the onset of eruption, where features
begin to move across the pixels more quickly, limiting the allowed
integration time.
For synthesizing the $V_{\lambda}$ and $I_{\lambda}$ profiles,
the LOS integrations are
first carried out at the resolution of the simulation domain.
The emergent line profiles for the individual plane-of-sky (POS)
grid points (at about 2 arcsec resolution) are then averaged over
the assumed observation reslution element ($\Delta x = 12$ arcsec).
These spatially averaged line profiles (at the
observation resolution) obtained for the individual simulation
snapshots (output at about 3 min intervals) are further averaged
over the integration time of about 12 min to obtain the final
line profiles $V_{\lambda}$ and $I_{\lambda}$ used to
infer $B_{\rm LOS}$ via equations (\ref{eqn:blos}), (\ref{eqn:V}), and
(\ref{eqn:I}), and $\sigma_B$ via equations (\ref{eqn:noise}) and
(\ref{eqn:photons}).
We further note that for the LOS integration, we have assumed that the
prominence condensations are optically thick, such that when the LOS hits
the cool plasma below temperature of $10^5 $K, we stop the integration
beyond that point, assuming the emission from the plasma behind is
obscured by the prominence and does not contribute to the final
emergent profile.

\section{Forward Modeled Results}
\begin{figure}[htb!]
\centering
\includegraphics[width=0.8\textwidth]{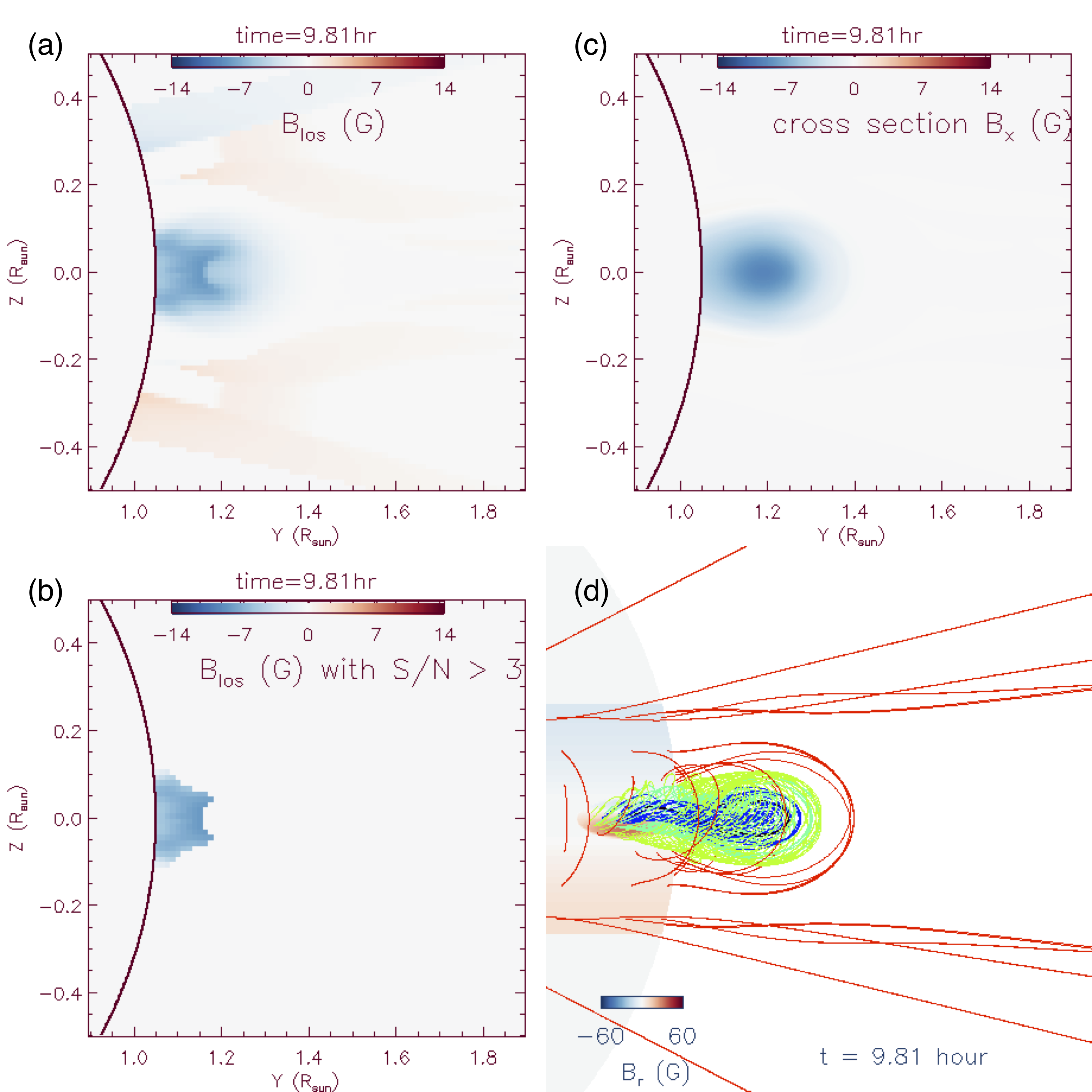}
\caption{(a) The inferred $B_{\rm LOS}$ (where the $x$-direction is
the LOS direction)
obtained from the forward synthesis
of the MHD model data (as described in Section \ref{sec:model}) at a time
during the quasi-static phase of the flux rope;
(b) The same as (a) but only shows the pixels of the measurable
$B_{\rm LOS}$ (see text); (c) The axial field
strength in the mid cross-section of the flux rope; (d) The 3D field lines of
the coronal magnetic field as viewed from the observer's perspective, with
the LOS parallel to the axis of the emerging flux rope.
A corresponding movie showing the evolution of the field strength maps
and the 3D flux rope view is available in
the online version.}
\label{fig:fig_blos_fdl_th90phm90}
\end{figure}

\begin{figure}[htb!]
\centering
\includegraphics[width=0.5\textwidth]{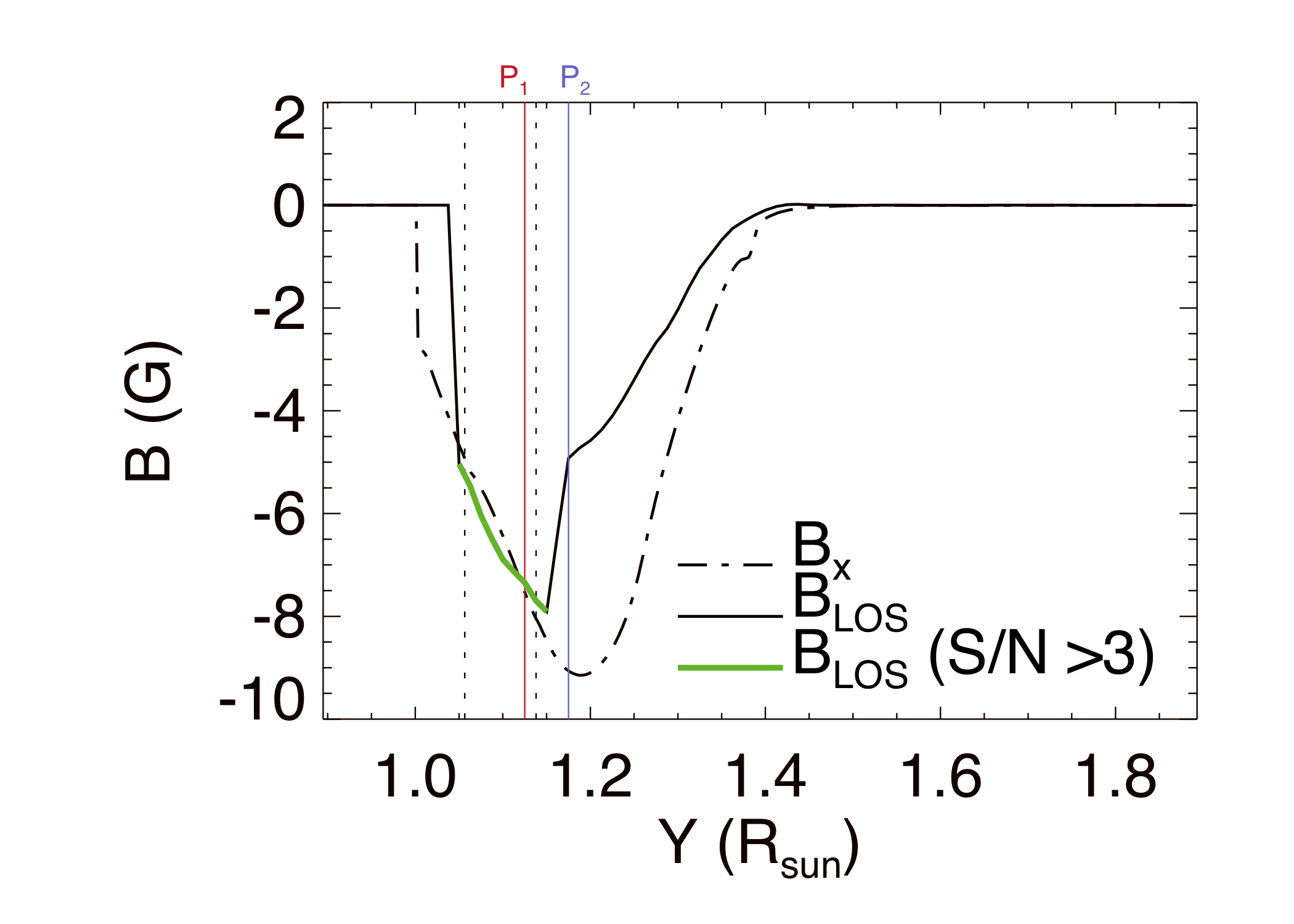}
\caption{Inferred $B_{\rm LOS}$ (solid line) along the line of
$Z=0.0063 R_{\rm sun}$ in the POS map shown in Figure \ref{fig:fig_blos_fdl_th90phm90}(a),
compared with the axial field $B_x$ (dashed-dotted line) along the same
$Z=0.0063 R_{\rm sun} $ line in the mid cross-section (also the POS cross-section) of
the flux rope shown in Figure \ref{fig:fig_blos_fdl_th90phm90}(c).The green solid
curve shows the portion of the measurable $B_{\rm LOS}$.
The dotted lines bracket the height range of the prominence. The
positions $P_1$ and $P_2$ indicated by the red and blue lines are referred to
later when discussing Figure \ref{fig:fig_p1p2}.}
\label{fig:fig_heightprof}
\end{figure}

\begin{figure}[htb!]
\centering
\includegraphics[width=0.45\textwidth]{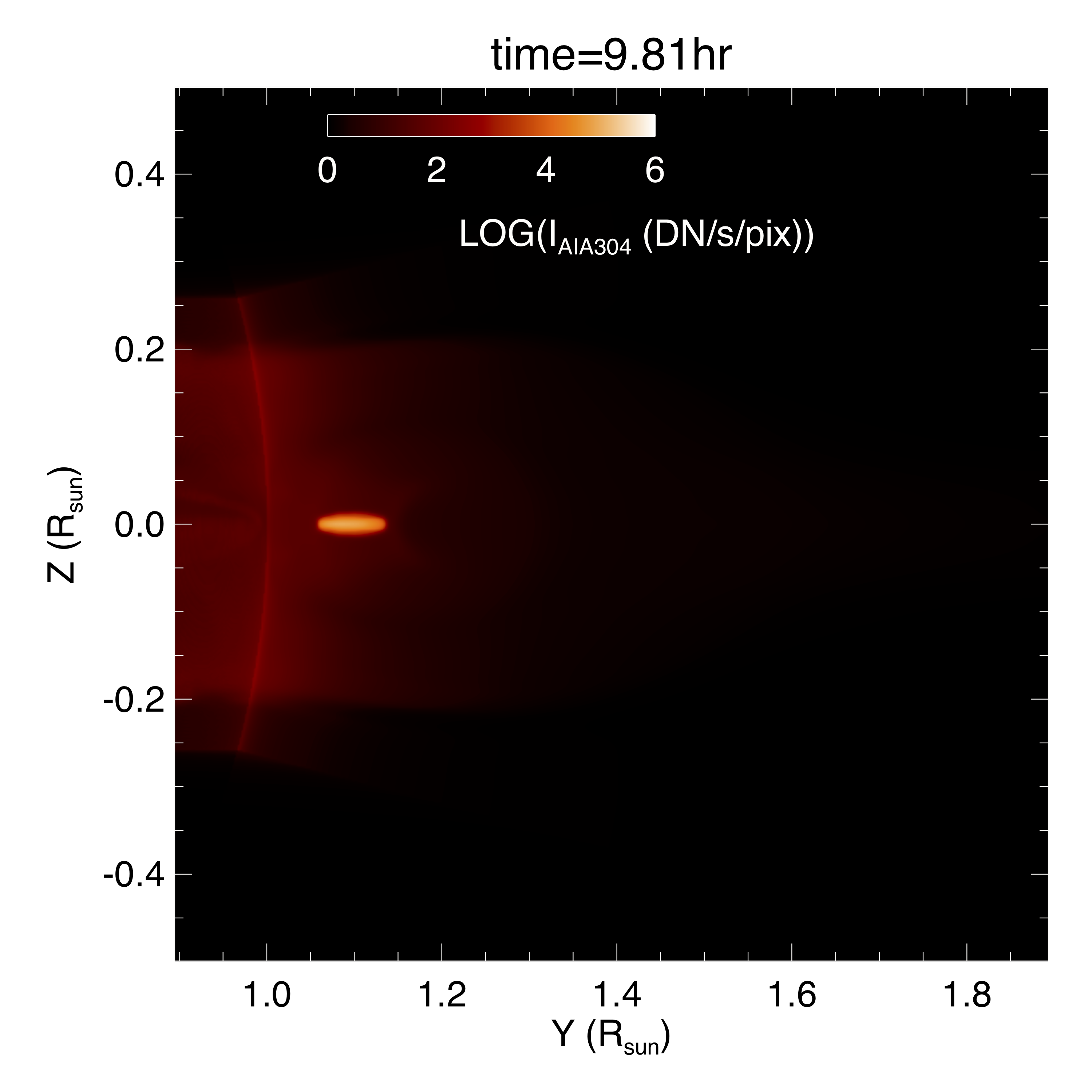}
\caption{Synthetic SDO/AIA 304 {\AA} channel emission from the same
viewing direction as that for the flux rope shown in
Figure \ref{fig:fig_blos_fdl_th90phm90}(d) at the
same time instant.}
\label{fig:fig_aia304_th90phm90}
\end{figure}

\begin{figure}[htb!]
\centering
\includegraphics[width=0.45\textwidth]{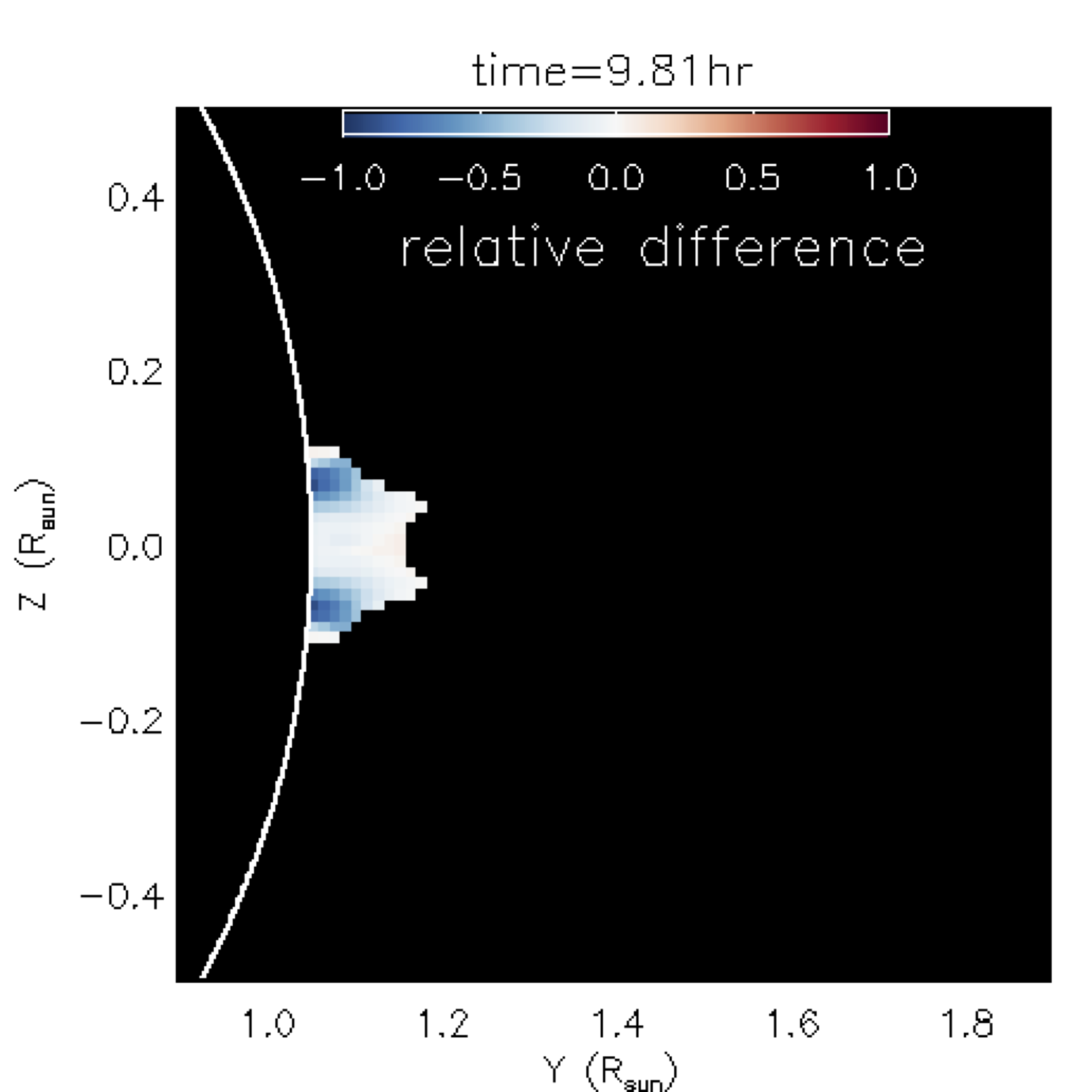}
\caption{Relative difference, $(B_{\rm LOS}-B_x)/|B_x|$, between the measurable
$B_{\rm LOS}$ shown in Figure \ref{fig:fig_blos_fdl_th90phm90}(b) and
the $B_x$ in the flux rope mid cross section shown in
Figure \ref{fig:fig_blos_fdl_th90phm90}(c).}
\label{fig:fig_diffblos}
\end{figure}

Figure \ref{fig:fig_blos_fdl_th90phm90}(a) shows a snapshot of the inferred
$B_{\rm LOS}$ from the synthesized V and I signals as described in the
above section, assuming the simulation domain is centered above the
west limb at the equator, with the LOS parallel to the axis of the
emerging flux rope.
Here the $x$-direction is the LOS direction pointing towards the
observer and the y-z plane denotes the plane-of-sky (POS).
Panel (b) is the same as (a) but only shows the inferred $B_{\rm LOS}$ for
which the signal-to-noise ratio ($| B_{\rm LOS} | / \sigma_B$) is above 3.
A $3 \sigma$ detection is typically considered a significant detection
with a 99.87 \% confidence that the result is real.
For the remainder of the paper, we call such inferred $B_{\rm LOS}$
as the {\it measurable} $B_{\rm LOS}$. 
For comparison, panel (c) shows the LOS component $B_x$ of the
magnetic field in the plane-of-sky cross-section (which corresponds to the
axial field in the mid cross-section of the flux rope), and panel (d) shows
the 3D magnetic field lines of the flux rope as viewed from the observer's
perspective.
It can be seen from the figure that one can detect a
measurable $B_{\rm LOS}$ that is of similar magnitude
(peaked at 7.9 G)
as the axial field of the flux rope in the mid cross section (peaked at
9.1 G), although the spatial profile of the inferred
$B_{\rm LOS}$ in the POS shows significant differences
from that of the axial field $B_x$ in the mid cross-section (at $x=0$) of
the flux rope.
To see this more quantitatively, Figure \ref{fig:fig_heightprof} shows
the inferred $B_{\rm LOS}$ as a function of height $Y$ along the
line near the center at $Z=0.0063 R_{\rm sun}$ in the POS map
shown in Figure \ref{fig:fig_blos_fdl_th90phm90}(a)) compared with the 
axial field $B_x$ along the same line at $Z=0.0063 R_{\rm sun} $ in the mid cross-section
of the flux rope shown in Figure \ref{fig:fig_blos_fdl_th90phm90}(c).
It can be seen that the two quantities are close for the range
from $Y=1.05 R_{\rm sun}$ (edge of the COSMO occulting disk) to about
$Y=1.14 R_{\rm sun}$, which is the height range of the prominence (see Figure
\ref{fig:fig_aia304_th90phm90}), above
which the magnitude of the inferred $B_{\rm LOS}$ becomes
significantly smaller than that of $B_x$.
Figure \ref{fig:fig_diffblos} shows the relative difference between the
measureable $B_{\rm LOS}$ in the POS (Figure
\ref{fig:fig_blos_fdl_th90phm90}(b)) and the axial field strength $B_x$
in the flux rope mid cross section (Figure \ref{fig:fig_blos_fdl_th90phm90}(c)).
It can be seen that the difference is small (less than 10\%) over the region
around the central prominence.
There are two regions further away from the prominence
(see the two blue blobs in the image),
where the measured $B_{\rm LOS}$ is significantly higher in magnitude (by about 90\%)
than the $B_x$ because the LOSs in these regions intersect the strong fields
in the legs of the flux rope in front of and behind the POS.

\begin{figure}[htb!]
\centering
\includegraphics[width=0.4\textwidth]{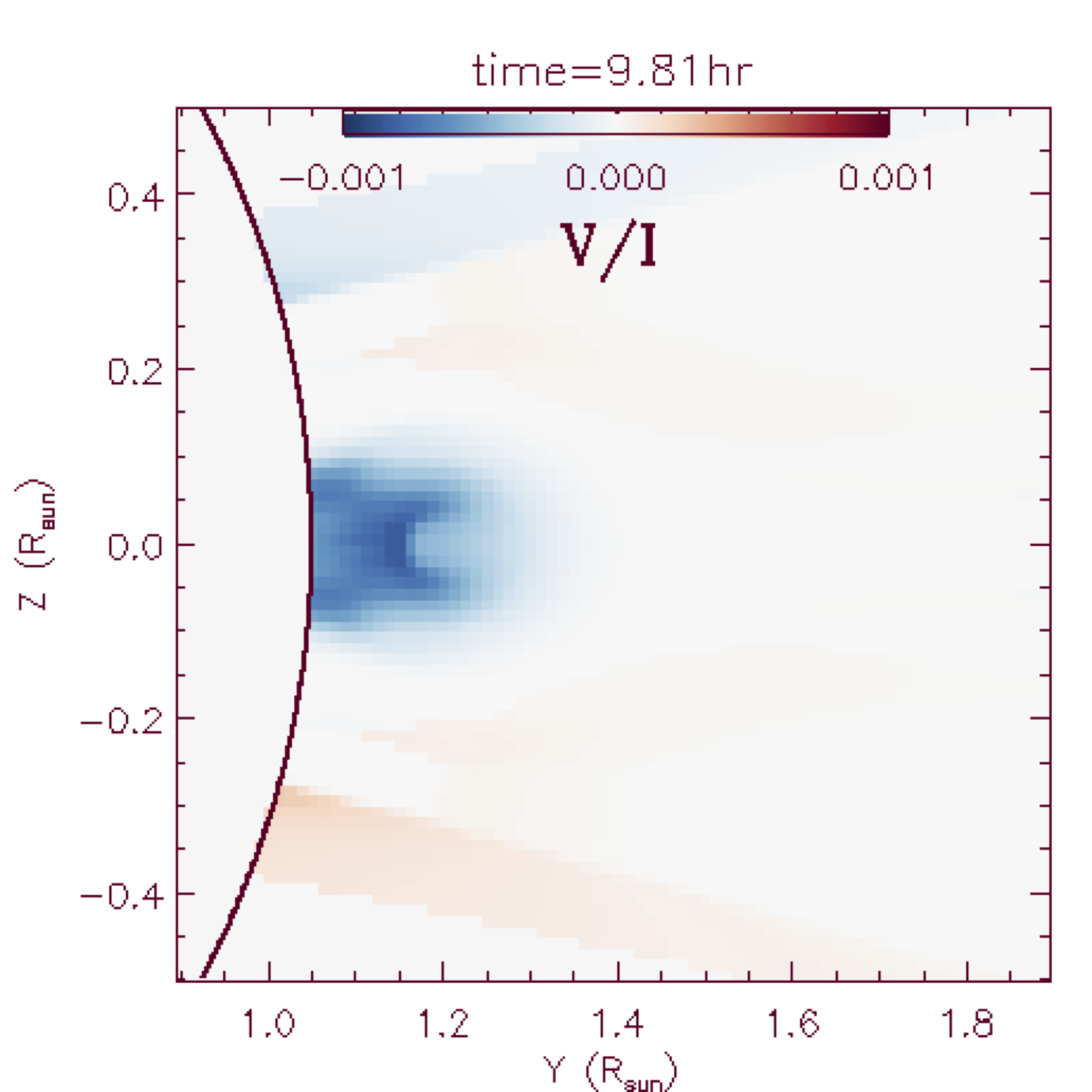}
\caption{The synthetic V/I signal corresponding to the same time
instance shown in Figure \ref{fig:fig_blos_fdl_th90phm90}.}
\label{fig:fig_VoI}
\end{figure}

\begin{figure}[htb!]
\centering
\includegraphics[width=0.5\textwidth]{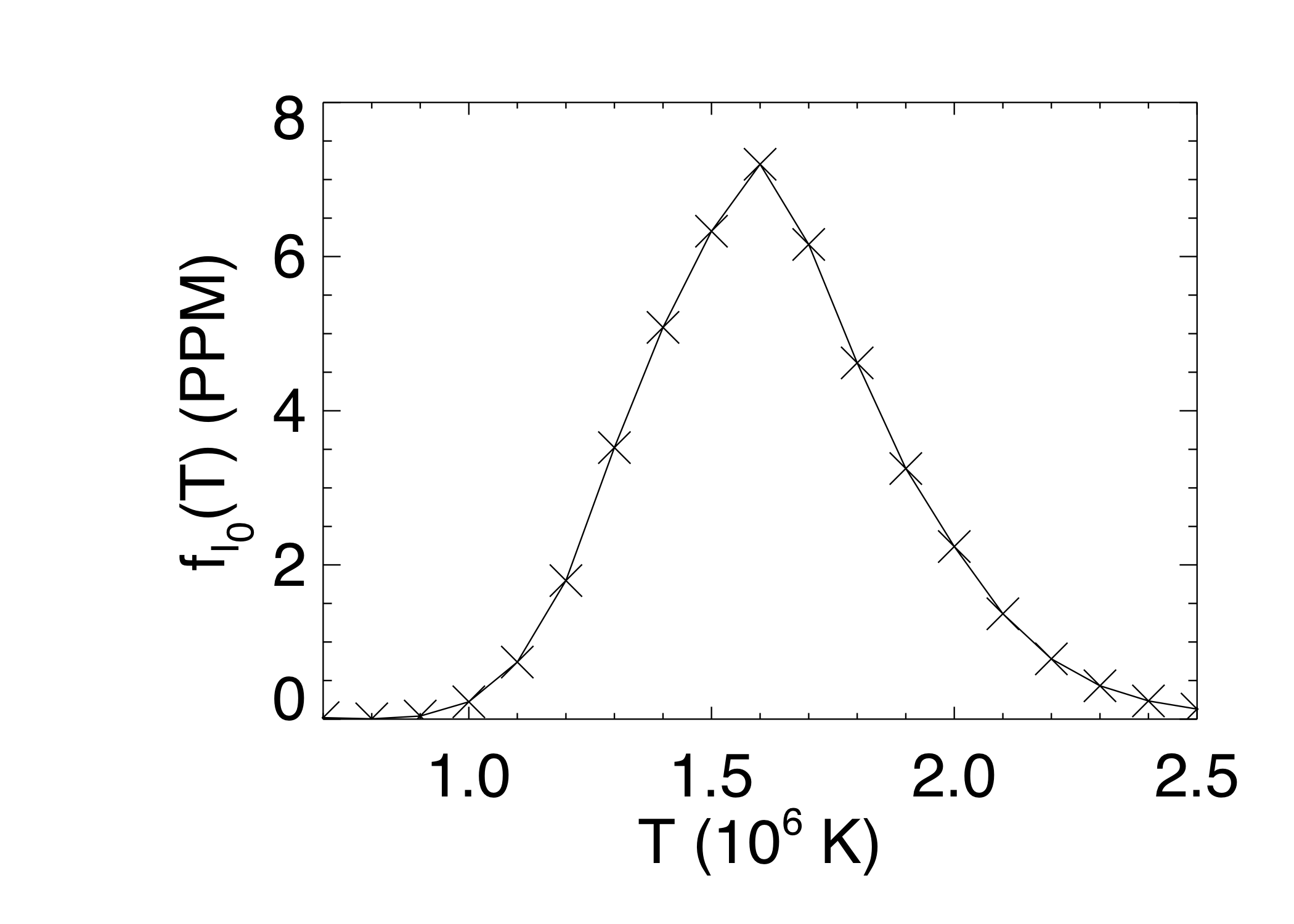}
\caption{The line center intensity ($f_{I_0}$) obtained from a LOS integration
with the CLE code assuming constant plasma properties (temperature, electron
density and magnetic field) along the LOS, as a function of the
temperature (T) value used for the LOS, while keeping the other plasma
parameters fixed (assuming electron density of $10^8 {\rm cm}^{-3}$, and
a constant magnetic field of 1G along the line of sight).}
\label{fig:fig_tempscan}
\end{figure}

\begin{figure}[htb!]
\centering
\includegraphics[width=0.4\textwidth]{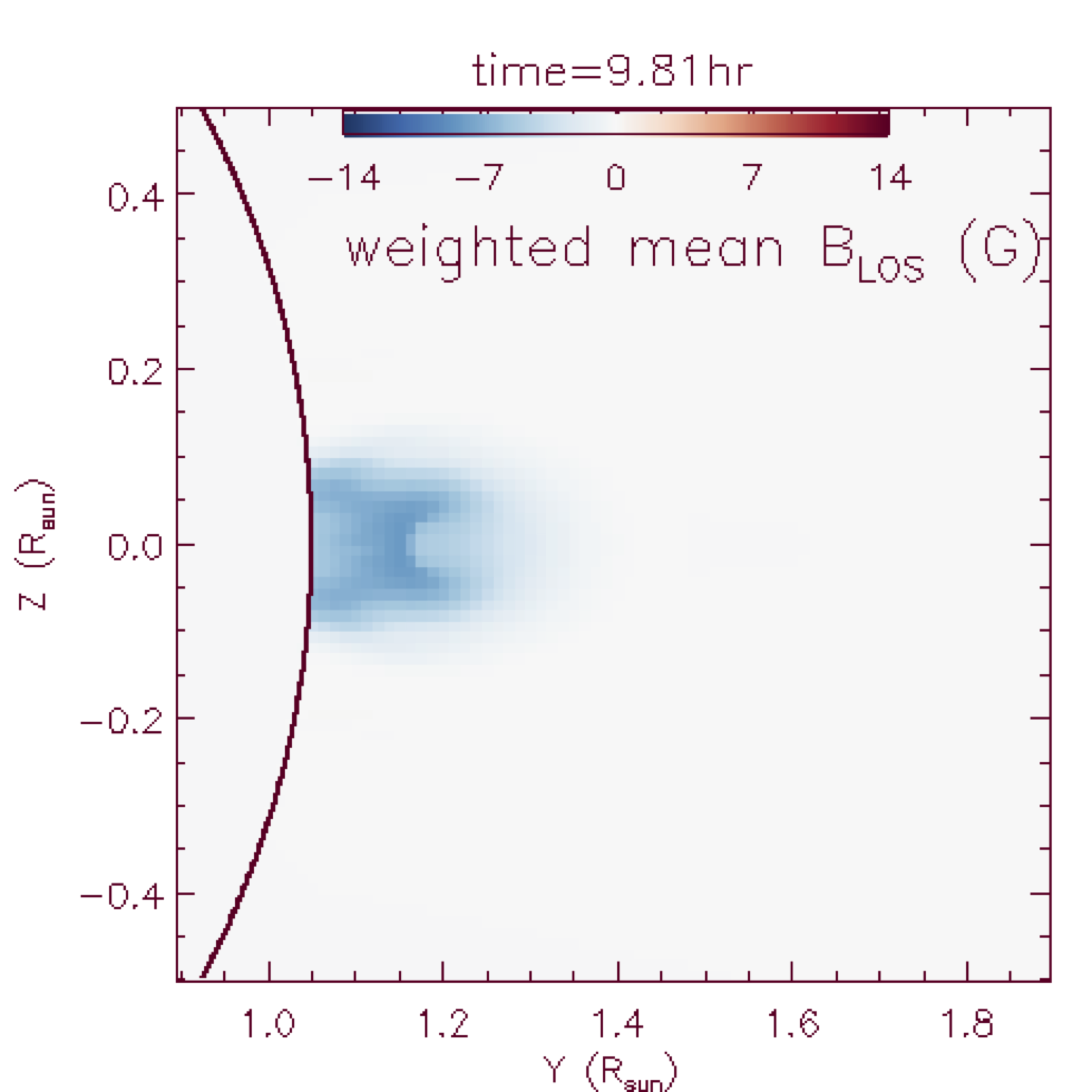}
\caption{POS map of the weighted-mean $B_{\rm LOS}$ over each LOS (see
text for how the weighted mean is computed).}
\label{fig:fig_blos_weightedmean}
\end{figure}

\begin{figure}[htb!]
\centering
\includegraphics[width=0.75\textwidth]{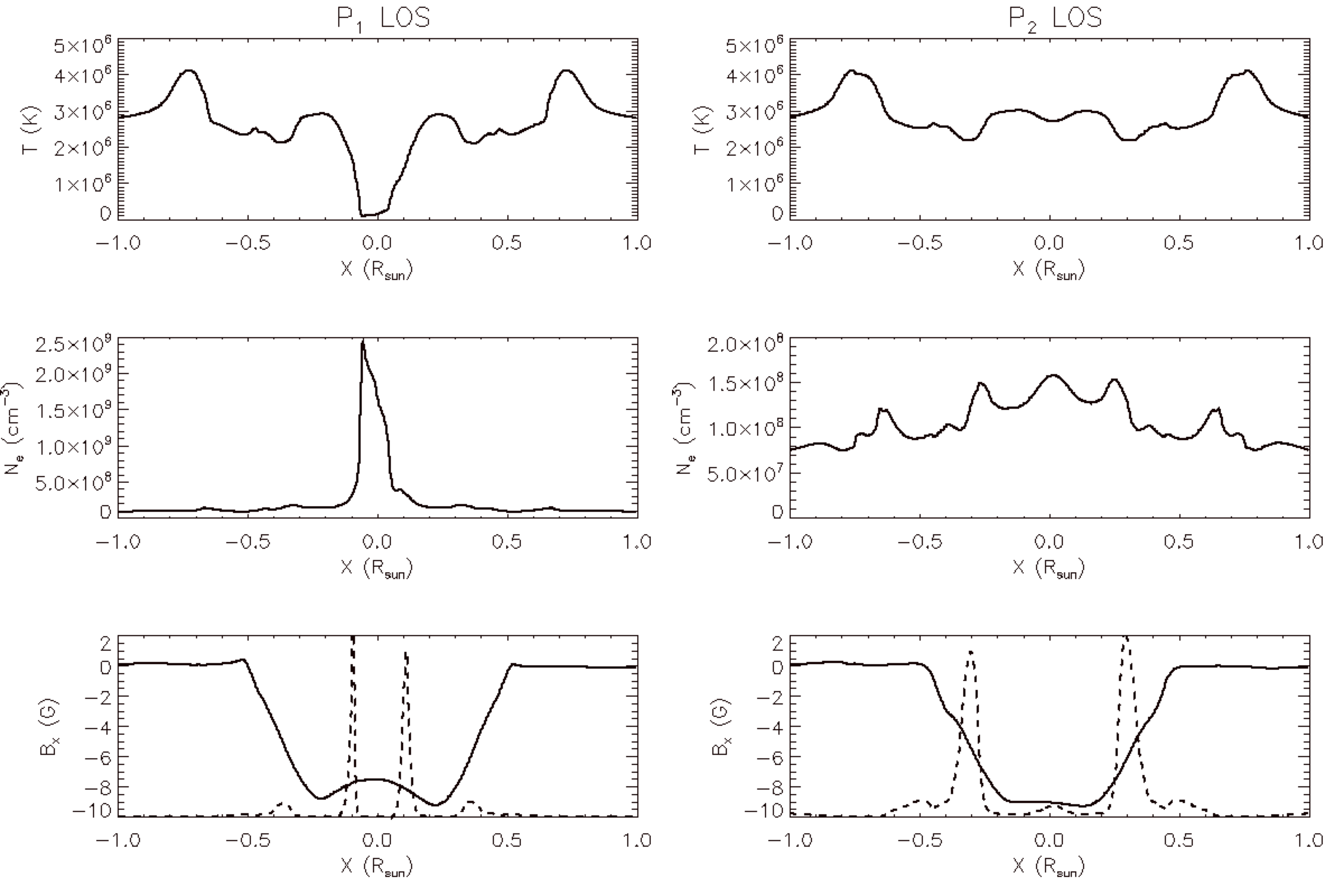}
\caption{Profiles of temperature T, electron density $N_e$, LOS field strength
$B_x$, along two example LOSs (left and right columns). The left (right)
column shows the LOS that produces the inferred $B_{\rm LOS}$ at
$P_1$ ($P_2$) marked by the red (blue) line position in
Figure \ref{fig:fig_heightprof}. The $P_1$ and
$P_2$ LOSs correspond to the POS positions of
($Y=1.127 R_{\rm sun}$, $Z=0.0063 R_{\rm sun}$) and ($Y=1.177 R_{\rm sun}$, $Z=0.0063 R_{\rm sun}$)
respectively in Figures \ref{fig:fig_blos_fdl_th90phm90}(a)(b).
The dashed line in the bottom panels shows the variation of the weight function
$W(x) \propto f_{I_0} (T(x)) N_e(x)$ along the LOS (x), in relation to the
variation of $B_x$ along the LOS.}
\label{fig:fig_p1p2}
\end{figure}

As given in equation (\ref{eqn:blos}), the inferred $B_{\rm LOS}$ is directly
proportional to the synthetic V/I (shown in Figure
\ref{fig:fig_VoI}),
with the proportionality factor varying with
the line width $\Gamma$. The synthetic V/I reaches a peak magnitude
of about 0.001 here for our modeled flux rope with $\sim 10$
G peak axial field strength.
As described in section \ref{sec:model}, the
V and I signals are given by the synthesized Stokes V and I
profiles of the FeXIII emission line through LOS integrations with
the CLE code.  Thus the inferred $B_{\rm LOS}$ is approximately
an FeXIII line intensity weighted mean of the LOS component $B_x$ of
the magnetic field along the LOS that goes through different parts
of the flux rope.  To see the temperature sensitivity of the
the FeXIII line, we show in Figure \ref{fig:fig_tempscan}
the line center intensity ($f_{I_0}$) obtained from a LOS integration with the
CLE code assuming constant plasma properties (temperature, electron
density and magnetic field) along the LOS, as a function of the (constant)
temperature (T) value used for the LOS while keeping the other plasma
parameters fixed.
It shows that the FeXIII line has a fairly narrow temperature sensitivity,
which peaks at about $T=1.6$ MK, but reduces by an order of magnitude when
$T$ rises to about $2.2$ MK or decreases to about $1.1$ MK.
The CLE code models the line emission under the combined influence of
resonance scattering and particle collisions.  The line intensity shows
a dependence on electron density $N_e$ that is between linear and 
quadratic \citep[e.g.][]{Gibson:etal:2016}.
Assuming that the FeXIII line intensity is approximately
proportional to $f_{I_0}(T) \, N_e$, we have computed a weighted-mean
$B_{\rm LOS}$ over each LOS: $\int B_x \, W(x) \, dx / \int W(x) \, dx$,
where $W(x) \propto f_{I_0} (T(x)) N_e(x)$ and x is the coordinate along
the LOS direction.  The resulting weighted-mean $B_{\rm LOS}$ in the POS
is shown in Figure \ref{fig:fig_blos_weightedmean}.
It is in good agreement with the inferred $B_{\rm LOS}$ shown in
Figure \ref{fig:fig_blos_fdl_th90phm90}(a) obtained from
the synthesized V and I profiles from the CLE code, in the region
of the flux rope.
Thus $W(x)$ gives a good approximation of the relative weight each part
of the LOS contributes to the inferred $B_{\rm LOS}$.
Figure \ref{fig:fig_p1p2} shows the profiles of various
quantities along two example LOSs from which the $B_{\rm LOS}$
at the $P_1$ and $P_2$ positions (marked by the red and blue lines)
in Figure \ref{fig:fig_heightprof}
are obtained.
For the $P_1$ LOS profiles shown in the left column panels, the LOS approaches the
prominence at the middle (x=0) of the LOS. The estimated weighting
function $W(x)$ (dashed line) along the LOS has two narrow peaks just
outside the prominence condensation, where T transitions steeply
from below $10^5$ K to the hot ``cavity'' temperature of above 2 MK,
going through the peak sensitivity temperature of 1.6 MK.  
Thus the inferred $B_{\rm LOS}$ is sampling the prominence carrying
field lines at the positions of prominence-to-cavity transition (further
illustration with Figure \ref{fig:promcfdl_W_temp} below), and
is fairly close to the axial field strength $B_x$ at
$x=0$ for the $P_1$ LOS
(see bottom left panel of Figure \ref{fig:fig_p1p2}).
Note for this LOS, because of the optically thick assumption
for the prominence condensations, only the part along the LOS from $x=1$
to $x=-0.07$ contributes to the LOS integration for synthesizing the
stokes profiles with the CLE code and also for evaluating the
weighted mean $B_{\rm LOS}$ shown in
Figure \ref{fig:fig_blos_weightedmean}.
We find that for all the LOSs that intersect the prominence 
vicinity, because
of the sampling property described above, the inferred $B_{\rm LOS}$
is close to the axial field $B_x$ at the central prominence dip at that
height (see the height range of $Y=1.05 R_{\rm sun}$ to $1.14 R_{\rm sun}$ in
Figure \ref{fig:fig_heightprof}).
Thus over this height range, the inferred
$B_{\rm LOS}$ can discern (with sufficient signal to noise ratio)
the increase with height of the
field strength of the prominence carrying field lines, which
is an indication that field lines are dipped
\citep[e.g. F17, section 1.8 in][]{Priest:2014}.
On the other hand, for the higher LOS at $P_2$
(see right column panels of Figure \ref{fig:fig_p1p2}),
the LOS is intersecting the hot core of the flux
rope above the prominence, with $T \approx 3$ MK around $x=0$ on
the LOS, where
$B_x$ is the strongest. The peak sampling represented by $W(x)$
(dashed line) is outside of the peak core field of the flux rope
(bottom right panel of Figure \ref{fig:fig_p1p2}),
in the region where the temperature is relatively lower outside
of the central hot core (top right panel of Figure \ref{fig:fig_p1p2}).
Thus the resulting inferred $B_{\rm LOS}$
is significantly lower than the mid cross-section axial field
$B_x$ at the $P_2$ height
as seen in Figure \ref{fig:fig_heightprof}.
Furthermore, because the temperature along the entire LOS for $P_2$ is high,
out of the sensitive range of the FeXIII line, the line intensity
obtained for this LOS is a factor of 10 smaller than
that of the $P_1$ LOS. The resulting inferred $B_{\rm LOS}$
at $P_2$ is thus below the error estimated from the photon noise, and
thus not in the measurable $B_{\rm LOS}$ map
in Figure \ref{fig:fig_blos_fdl_th90phm90}(b).

\begin{figure}[htb!]
\centering
\includegraphics[width=0.8\textwidth]{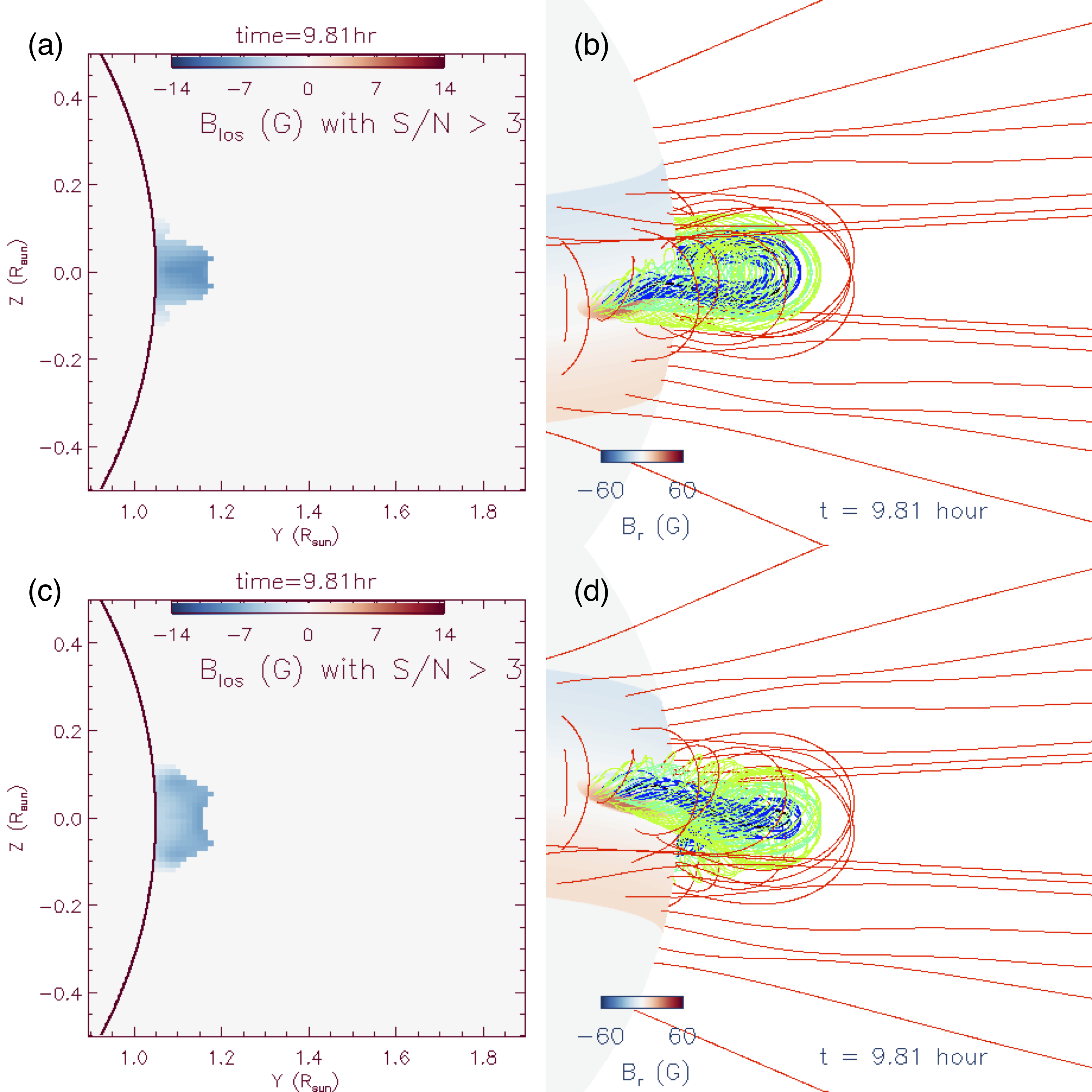}
\caption{The measurable $B_{\rm LOS}$ (left panels)
obtained from the forward synthesis of the MHD model data (as described
in Section \ref{sec:model}) for two different viewing angles as illustrated
by the 3D field line plots (right panels), with the flux rope centered
above the limb as in
Figure \ref{fig:fig_blos_fdl_th90phm90} but rotated
from the azimuthal direction by $7^{\circ}$ anti-clockwise (upper panels) and
clockwise (lower panels) respectively at the same time instant during the
quasi-static phase of the flux rope.}
\label{fig:fig_blos_fdl_rot}
\end{figure}

\begin{figure}[htb!]
\centering
\includegraphics[width=0.5\textwidth]{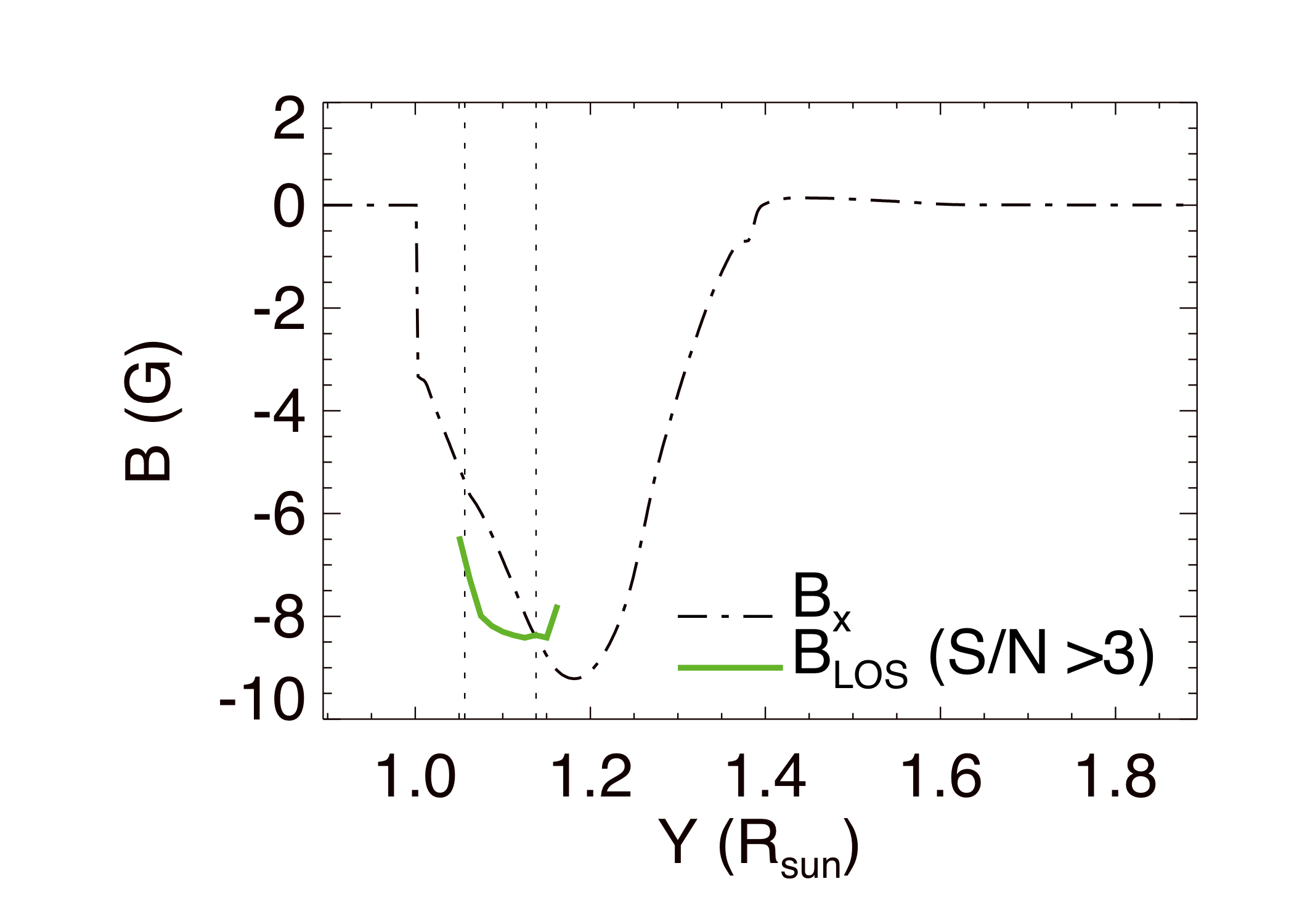} \\
\includegraphics[width=0.5\textwidth]{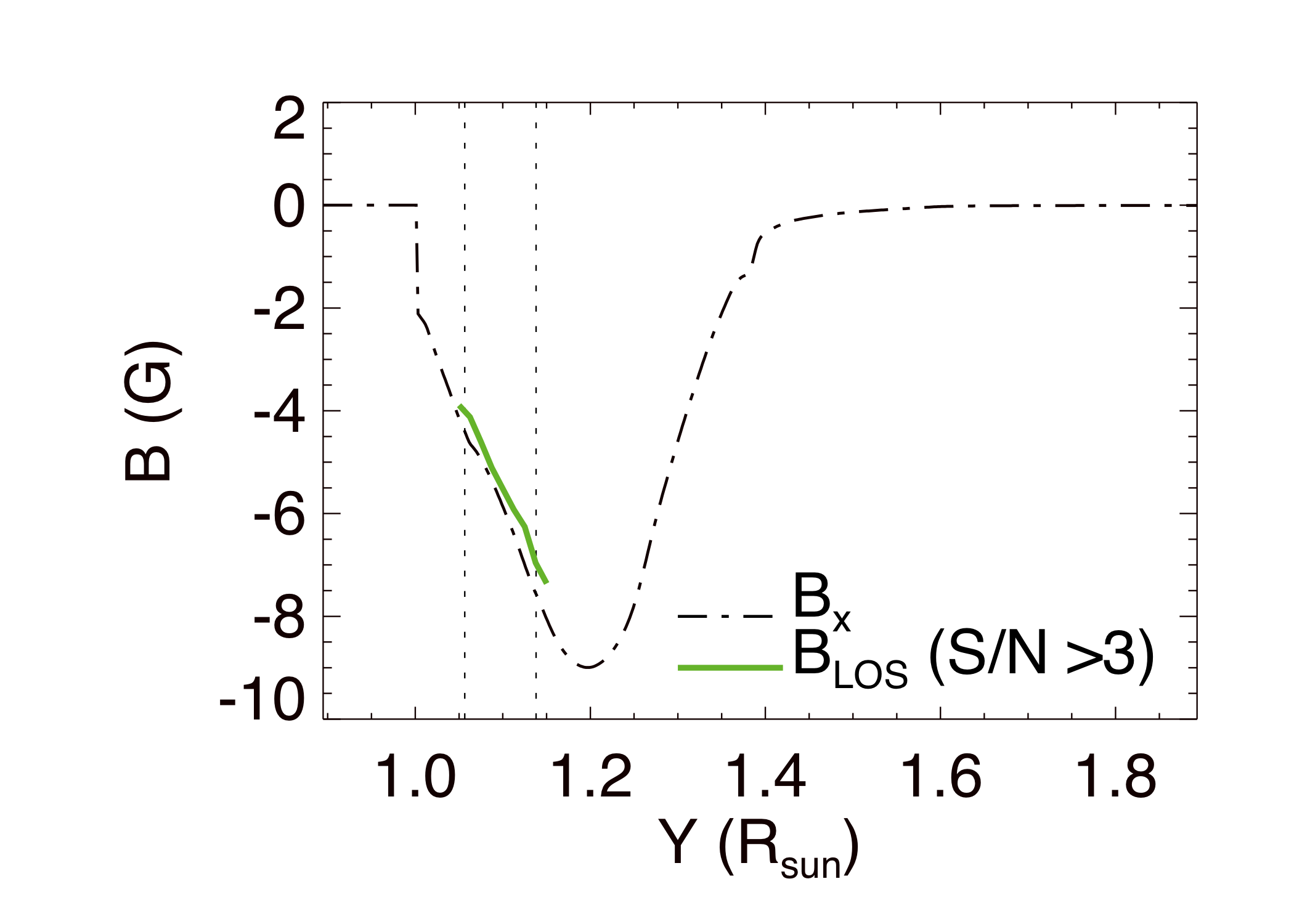}
\caption{Profile of the measurable
$B_{\rm LOS}$ (green curves) as a function of height $Y$ along
the line of $Z=0.0063 R_{\rm sun}$ in the POS map
shown in Figure \ref{fig:fig_blos_fdl_rot}(a) (upper panel) and the POS map
shown in Figure \ref{fig:fig_blos_fdl_rot}(c) (lower panel). For comparison
the LOS field $B_x$ along the same line of $Z=0.0063 R_{\rm sun}$ in the
POS cross-section of the flux rope is shown as the dash-dotted lines.}
\label{fig:fig_heightprof_rot}
\end{figure}

Figure \ref{fig:fig_blos_fdl_rot} shows the
measurable $B_{\rm LOS}$ (left panels) for two other viewing
angles, where the flux rope centered above the limb (see the right
panels for the 3D views in
Figure \ref{fig:fig_blos_fdl_th90phm90}) is rotated from the
azimuthal direction by $7^{\circ}$ anti-clockwise (upper panels) and
clockwise (lower panels) respectively. 
For both viewing angles, one can measure a significant $B_{\rm LOS}$ (with
a peak value of 8.4 G and 7.4 G respectively) in the flux rope region in
the POS over the height range of the prominence, similar to the results
for the previous viewing angle shown in Figure
\ref{fig:fig_blos_fdl_th90phm90}(b)(d).
In all three cases (see the POS maps in Figure \ref{fig:fig_blos_fdl_th90phm90}(b) and
Figures \ref{fig:fig_blos_fdl_rot}(a)(c)),
the measurable $B_{\rm LOS}$ shows an increase with height 
(Y) in the central region near $Z=0$ (near the prominence).
This is shown more quantitatively
by the green solid curves in
Figure \ref{fig:fig_heightprof_rot} and Figure
\ref{fig:fig_heightprof}, which show the profiles of the
measurable $B_{\rm LOS}$
along the line of $Z=0.0063 R_{\rm sun}$ in the POS maps. The prominence
height range is between the dotted lines.
The measurable $B_{\rm LOS}$ profiles all show increase with
height over the prominence height range and their values are fairly close
to the LOS field strengths of the flux rope in the mid cross-section.

The reason for this is further illustrated in
Figure \ref{fig:promcfdl_W_temp}.
\begin{figure}[htb!]
\centering
\includegraphics[width=0.7\textwidth]{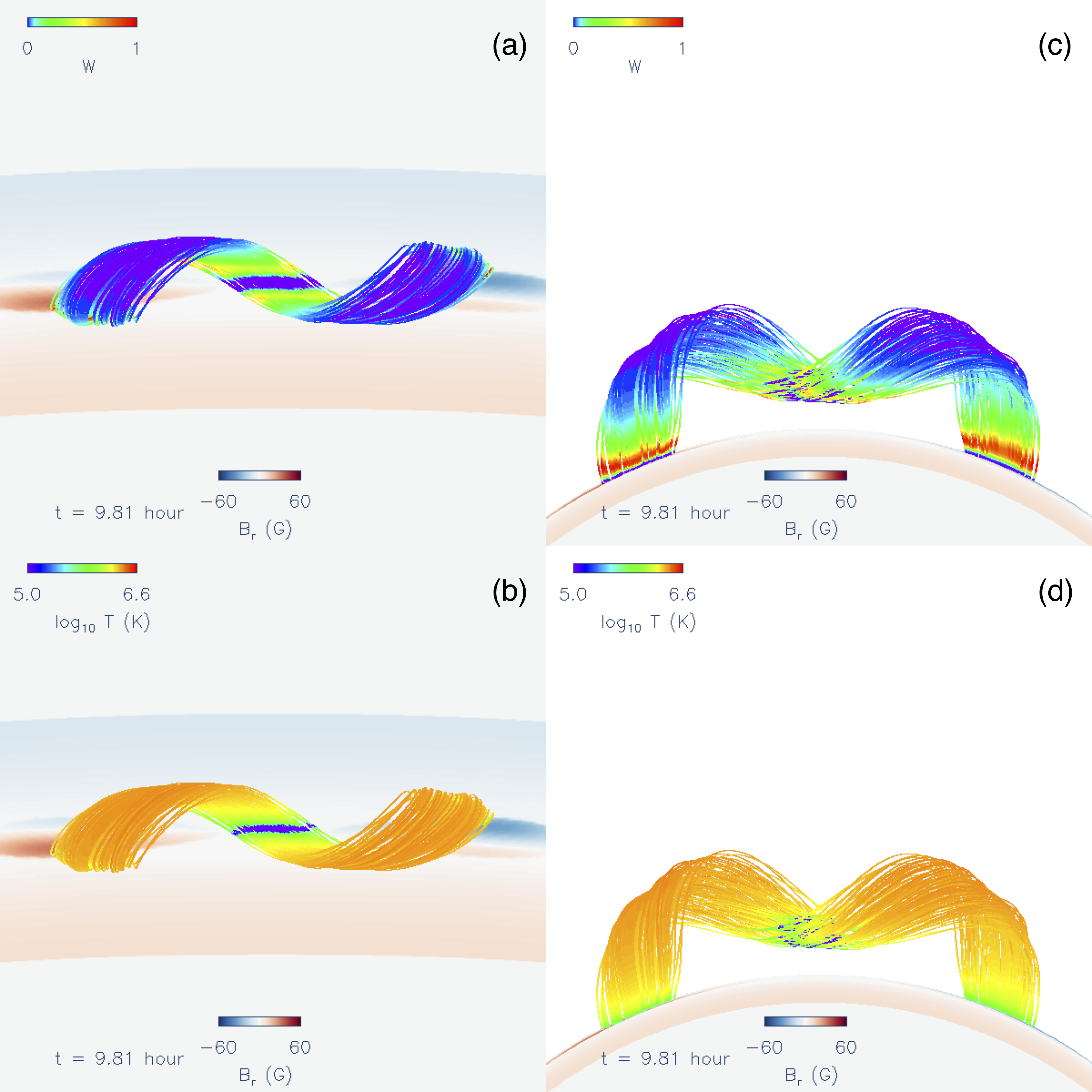}
\caption{3D views of the densely traced prominence carrying field lines
colored based on the weighting function
$W \propto f_{I_0} (T) N_e$ (top panels),
and corresponding views of the field lines colored based on
the temperature (bottom panels)}
\label{fig:promcfdl_W_temp}
\end{figure}
The top panels show 3D views of the densely traced
prominence carrying field lines colored with the weighting function 
$W \propto f_{I_0} (T) N_e$, showing which parts of the field lines
contribute strongly (with higher $W$) to the FeXIII emission and hence to the
$B_{\rm LOS}$ measurement.  For reference the temperature distribution
along the prominence field lines are shown in
the lower panels of Figure \ref{fig:promcfdl_W_temp}.
It can be seen that the parts that the
$B_{\rm LOS}$ measurement is sensitive to are just outside
of the central prominence condensations, at the transition from prominence
temperature to the hot cavity temperature along the prominence carrying
field lines. This further illustrates that {\it the measured $B_{\rm LOS}$
for the prominence height range
tends to be fairly close to the field strength at the prominence dip
and reflects its variation (increase) with height}.
We emphasize that this localization of the measured field strength is
because of the specific temperature configuration in the simulation that
the signal comes mostly from around the cold prominence. If the temperature
distribution in the flux rope outside of the prominence is more uniformly
close to the 1.6 MK formation temperature of the FeXIII line, then the
measured field strength would not reflect so closely the field strength
of the prominence dips, as we will show later in a different case at
the end of this section.

\begin{figure}[htb!]
\centering
\includegraphics[width=0.8\textwidth]{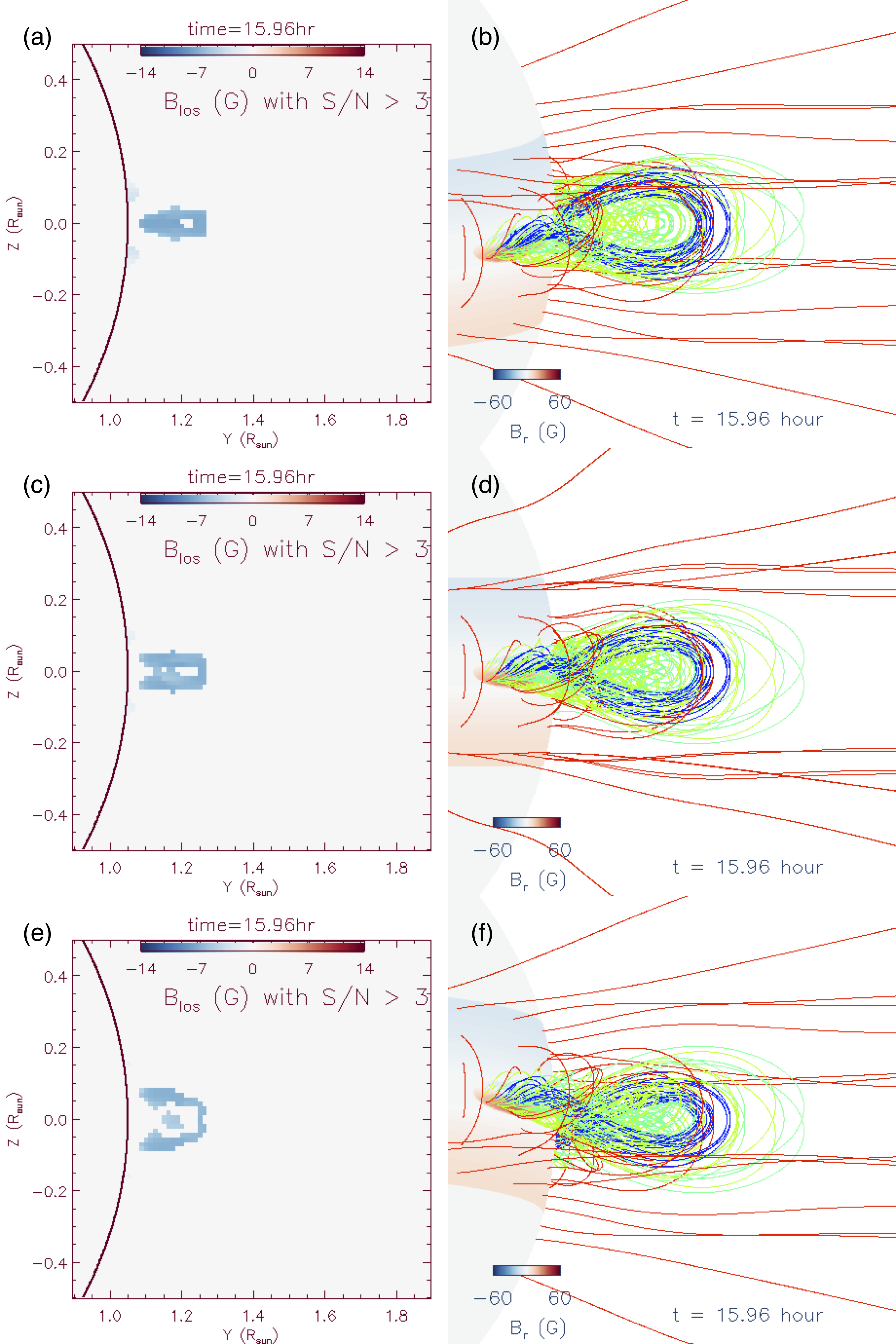}
\caption{The top, middle, and bottom rows are the same as
Figures \ref{fig:fig_blos_fdl_rot}(a)(b),
Figures \ref{fig:fig_blos_fdl_th90phm90}(b)(d),
and Figures \ref{fig:fig_blos_fdl_rot}(c)(d) respectively,
except at a later time instance towards the end of the slow rise phase.
A corresponding
movie of the evolution of the measurable $B_{\rm LOS}$ maps and the flux rope views
is also available in the online version.}
\label{fig:fig_blos_fdl_evol}
\end{figure}

Figure \ref{fig:fig_blos_fdl_evol} shows the POS maps of the
measurable $B_{\rm LOS}$ and the 3D flux rope viewed from the
corresponding observation perspectives at a later time instance
towards the end of the slow rise phase, just before the onset of eruption.
It corresponds to the time instance shown in Figure
\ref{fig:fig_fdl_aia304_ev}(b)(e), where the flux rope has become
significantly kinked and the prominence is rising.
A movie of the evolution of the $B_{\rm LOS}$ maps and the
flux rope views corresponding to Figure \ref{fig:fig_blos_fdl_evol} is
also available in the online version.
We can see from the Figure and the movie that we can detect a
measurable out-moving $B_{\rm LOS}$ through the slow
rise phase up to a height of about $y=1.3 R_{\rm sun}$, and until a time
when the flux rope has accelerated to a speed of
about $41$ km/s.  After that, we no longer detect a measurable outward
moving field.

\begin{figure}[htb!]
\centering
\includegraphics[width=0.8\textwidth]{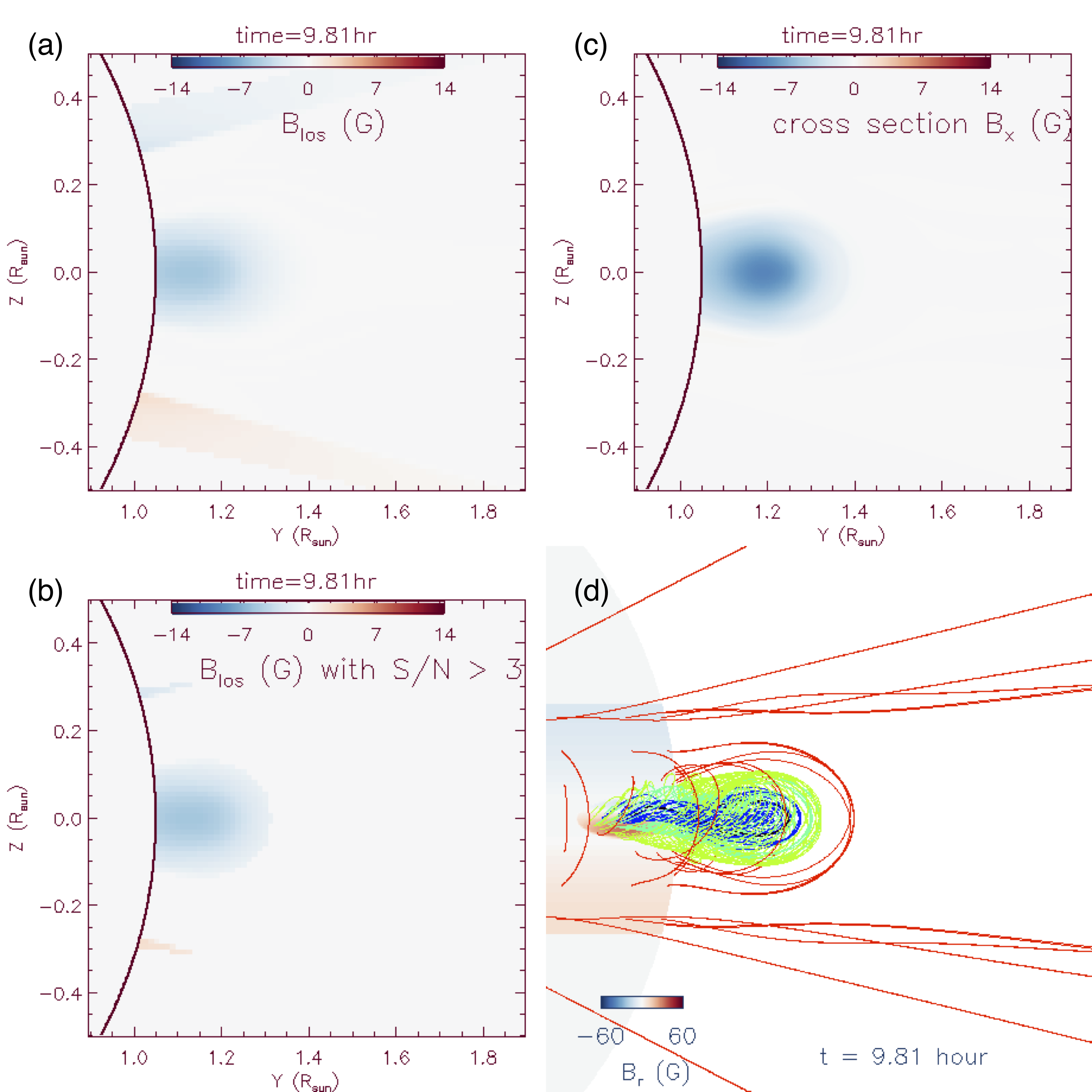}
\caption{Same as Figure \ref{fig:fig_blos_fdl_th90phm90} except using
the hysrostatic isothermal corona at 1.6 MK to replace the
plasma properties in the simulation domain.}
\label{fig:fig_blos_fdl_th90phm90_iso_it198}
\end{figure}

\begin{figure}[htb!]
\centering
\includegraphics[width=0.5\textwidth]{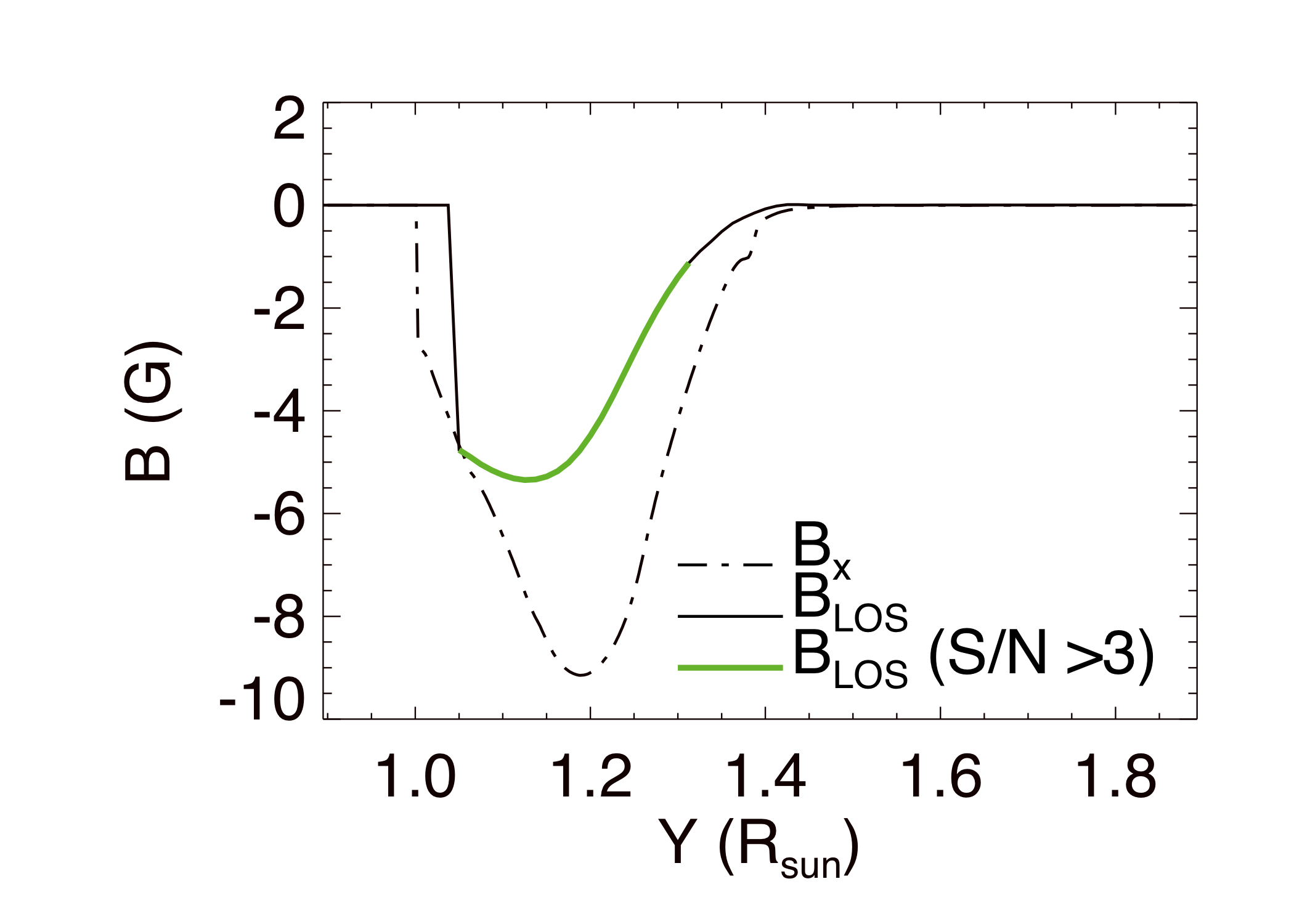}
\caption{Same as Figure \ref{fig:fig_heightprof} except for the case with
the hysrostatic isothermal corona at 1.6 MK 
replacing the plasma properties in the simulation domain.}
\label{fig:fig_heightprof_iso}
\end{figure}

\begin{figure}[htb!]
\centering
\includegraphics[width=0.8\textwidth]{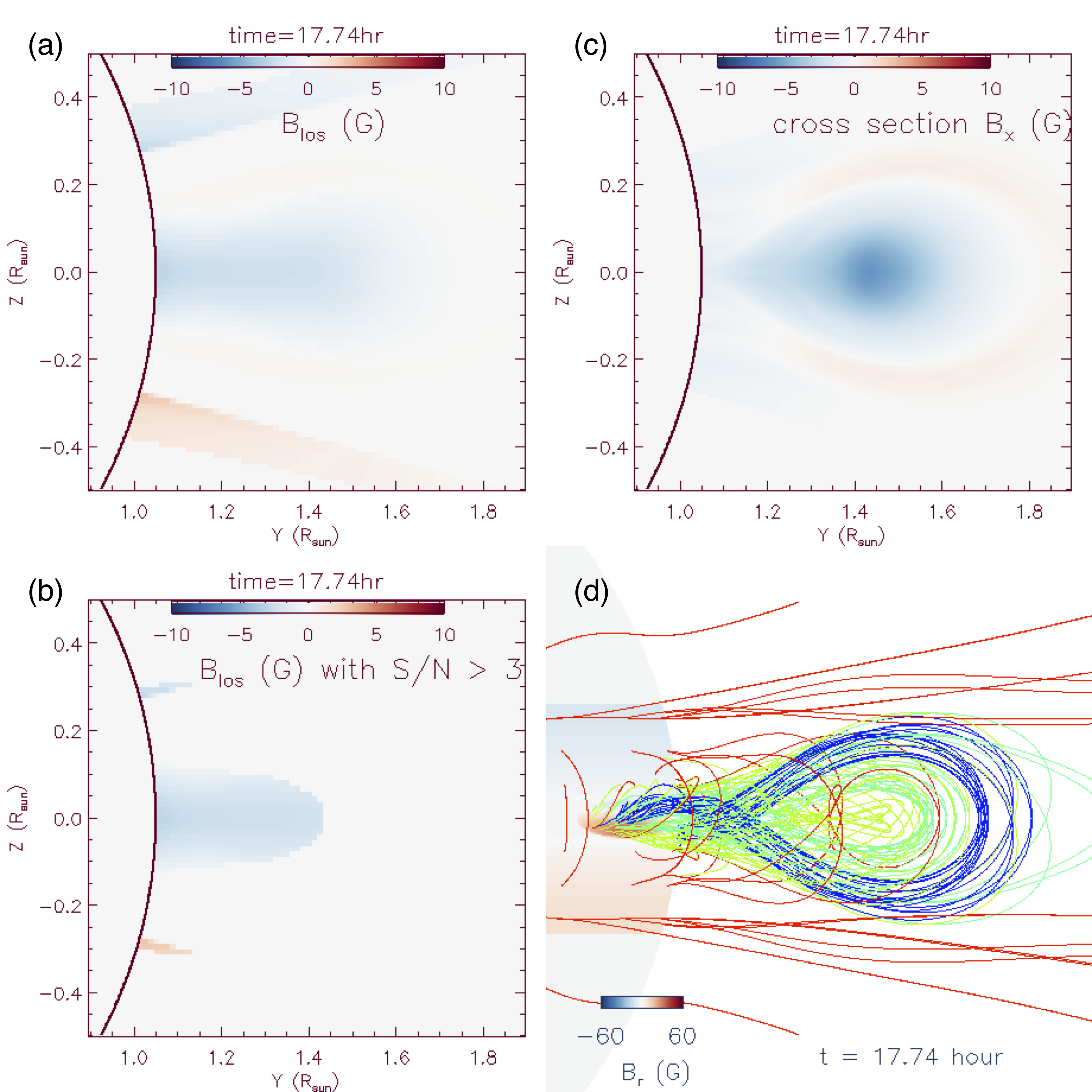}
\caption{Same as Figure \ref{fig:fig_blos_fdl_th90phm90_iso_it198} except
for a later time at $t=17.74$ hr when the flux rope has begun to erupt, and
also the change in the color table range for the field strength.
A corresponding movie showing the evolution of the magnetic field is
also available in the online version.}
\label{fig:fig_blos_fdl_th90phm90_iso_it358}
\end{figure}

There are significant uncertainties in regard to the plasma thermodynamic
properties obtained in the simulation, for which a highly simplified coronal
heating is assumed. We therefore have considered an alternative extreme
case where we replace the plasma properties (temperature, density, and velocity)
in the simulation domain with a hydrostatic isothermal atmosphere with
$T= 1.6$ MK (at maximum sensitivity of the FeXIII line) and a base
density of $5 \times 10^8 {\rm cm}^{-3}$, only keeping
the simulated magnetic field evolution for the forward modeling to
see how much the resulting inference of the $B_{\rm LOS}$ can differ.
The results for the inferred and measurable $B_{\rm LOS}$ for the same
observation view as that of Figure \ref{fig:fig_blos_fdl_th90phm90} are
shown in Figure \ref{fig:fig_blos_fdl_th90phm90_iso_it198}.
Comparing this to Figure \ref{fig:fig_blos_fdl_th90phm90},
we find that the measurement sensitivity is significantly increased such
that the observation can now measure a significant $B_{\rm LOS}$ nearly
throughout the flux rope cross section in the POS
(Figure \ref{fig:fig_blos_fdl_th90phm90_iso_it198}(b)),
although the magnitude of the measured field strength
significantly under estimates the axial field strength $B_x$ in the
mid cross section of the flux rope
(Figure \ref{fig:fig_blos_fdl_th90phm90_iso_it198}(c)).
Figure \ref{fig:fig_heightprof_iso} shows a quantitative comparison
of the field strength profiles along the central slice at
$Z=0.0063 R_{\rm sun}$ in the POS.
The measured peak field strength of $B_{\rm LOS}$
is roughly a half of the peak axial field strength $B_x$ of the flux rope.  
The under estimate of the flux rope field strength by the measured
$B_{\rm LOS}$ is due to the broad averaging along the LOSs with much
more uniform weighting (because of the uniform temperature and smoother
density variation) through the flux rope and the outside
arcade field where the $B_{\rm LOS}$ component is weak.
Nevertheless, the measured $B_{\rm LOS}$ still shows a region of (weaker) increase
of field strengt with height in the lower height range of the flux rope
cross section, indicative of the concave upturning field geometry of
the flux rope there.
Figure \ref{fig:fig_blos_fdl_th90phm90_iso_it358} shows the measured
$B_{\rm LOS}$ in the POS at a later time when the flux rope has begun
to erupt, and a movie corresponding to the Figure showing the temporal
evolution of the measured $B_{\rm LOS}$.
We find that in this case we can detect a measurable outgoing $B_{\rm LOS}$
to a greater height (about $1.43 R_{\rm sun}$) and over a significantly
larger area of the rising flux rope cross section, until a time when the
flux rope has accelerated to a speed of $485$ km/s.
   
\section{Discussion and Conclusion}
Using an MHD simulation of a prominence carrying coronal
flux rope, we have carried out forward synthesis of the circular
polarization measurement by the proposed COSMO LC to examine its
capability of inferring the LOS magnetic field of the flux rope
above the limb, viewed nearly along the length of the flux rope.
We found that using an integration time of about 12 min and an
observational resolution of 12 arcsec, the LC can measure
with sufficient signal to noise ratio ($>3$) the
flux rope field strength of a few G in the
region around the prominence,
within the height range of the prominence.
We find that the inferred $B_{\rm LOS}$ can be fairly well approximated
by a mean of the LOS component of ${\bf B}$ along the LOS with
a weighting that is proportional to $f_{I_0} (T) N_e$,
where $f_{I_0} (T)$ describes the
temperature sensitivity of the FeXIII line intensity as shown in
Figure \ref{fig:fig_tempscan}, with a narrow peak at $1.6 $MK.
Because of this and the temperature configuration of the simulated
flux rope,
we find that
for those LOS that intersect the prominence vicinity,
the measured $B_{\rm LOS}$
is localized to the region of prominence-to-hot-cavity transition
and is fairly close to the axial field strength at the prominence
dip.
As a result the measurable $B_{\rm LOS}$ shows fairly accurately
its increase with height profile, which is a signature of the concave
upturning dipped field lines supporting the prominence.

Above the prominence heights, the flux rope develops a hot core
which reaches a peak temperature of about $3$ MK in our MHD model.
As a result the LOSs at these heights tend to sample the field strength
outside of the hot core strong field region (e.g. the P2 LOS shown
in right column of Figure \ref{fig:fig_p1p2}), and the inferred
$B_{\rm LOS}$ is significantly below the axial field at the corresponding
height in the mid cross section of the flux rope, and also does not have
enough signal to noise ratio to make it measurable.
It is likely that our MHD model is over-estimating the temperature
in the hot cavity in the flux rope surrounding the prominence
condensation due to the simplified empirical coronal heating used and
also due to the heating from the numerical diffusion of the magnetic
field. This may have unrealistically reduced the sensitivity of the
FeXIII emission line measurement for the flux rope cavity region. 
We have therefore also examined a case where we have replaced the
plasma properties in the simulation domain with a hydrostatic
isothermal atmosphere at the peak sensitivity temperature of 1.6 MK
for the FeXIII line. In that case we found that the measurement
sensitivity is greatly improved such that a significant $B_{\rm LOS}$
can be measured throughout the flux rope cross section in the POS.
Although the magnitude of the measured field significantly under
estimates the axial field strength of the flux rope because of the
broad averaging along the LOS, nevertheless it still detects an
increase with height profile in the lower height range, indicating
the concave up turning field geometry of the flux rope there.
On the other hand X-ray observation of a filament cavity by
\citet{Hudson:etal:1999}
and eclipse observations by \citet{Habbal:etal:2010} have indicated
high temperatures, of about 2 MK or even higher, for the cavity
regions surrounding the prominences.
Thus it may be difficult to measure magnetic field strength in
the hot cavity region of the flux rope far away from the prominence itself with
the FeXIII line. 
However, our forward analysis shows that it is feasible to measure
the field strength in the vicinity of the prominence with the signal
from the region of prominence-to-cavity transition
as discussed above. This provides important information about the
properties of the prominence carrying magnetic fields.

Furthermore, we find that the synthetic observation can measure the outward
moving magnetic field around the rising prominence during the slow rise phase
as the flux rope develops the kink instability (F17). It can detect
the outward moving field up to a height of
about $1.3 R_{\rm sun}$, until a time when the flux rope accelerates to
about $41$ km/s, after which it can no longer detect a measurable field because
of the rapid change and the weakening of the field strength.
Thus our synthetic observation can track the field strength evolution
of the prominence supporting field into its early phase of the onset
of eruption.
Examination of the extreme case with the isothermal atmosphere with 1.6 MK shows
a greatly improved sensitivity in measuring the outward moving field during the
onset of the eruption. It can detect a measurable outgoing $B_{\rm LOS}$
over a larger area of the rising flux rope cross section (Figure
\ref{fig:fig_blos_fdl_th90phm90_iso_it358}),
to a greater height of about $1.43 R_{\rm sun}$ and
until a time when the flux rope has accelerated to a speed of $485$ km/s.
A further improved approach that combines measurements of multiple coronal
emission lines, with different temperature sensitivities, might allow a more
comprehensive picture of the magnetic field throughout the
multi-thermal
prominence-cavity system. We leave this for a future study.

\acknowledgments
We thank the anonymous referee for helpful comments that improved the paper.
We thank Roberto Casini for helping with using the CLE code, and for
reviewing the manuscript and helpful discussions.
We thank Haosheng Lin for helpful discussions and notes on calculating the noise.
This work is supported in part by the Air Force Office of Scientific Research
grant FA9550-15-1-0030 to NCAR. NCAR is sponsored by the National Science
Foundation. The numerical simulations were carried out on the Cheyenne
supercomputer at NWSC under the NCAR Strategic capability project NHAO0011 and
also on the DOD supercomputer Topaz at ERDC under the
project AFOSR4033B701.

\end{document}